\documentclass[10pt,twocolumns]{IEEEtran}
\usepackage{amsmath,amsfonts,amssymb,amsbsy, amsthm,ae,aecompl}
\usepackage{algorithm,algpseudocode}
\usepackage[utf8]{inputenc}
\usepackage[english]{babel}
\usepackage{bm}
\usepackage{color}
\usepackage[noadjust]{cite}
\usepackage{epsfig}
\usepackage{enumerate}
\usepackage{float} 
\usepackage{fancyhdr}
\usepackage[T1]{fontenc}
\usepackage[acronym,toc,shortcuts]{glossaries}
\usepackage{graphicx, caption, subcaption}
\usepackage{hyperref}
\usepackage{lastpage}
\usepackage{listings}
\usepackage{lipsum}
\usepackage{dsfont}
\usepackage{multind}
\usepackage[normalem]{ulem}
\usepackage{transparent}
\usepackage{tikz,pgfplots}
\usepackage{times}
\usepackage{verbatim}
\usepackage[all]{xy}

\usetikzlibrary{calc}

\hypersetup{
    colorlinks,%
    citecolor=black,%
    filecolor=black,%
    linkcolor=black,%
    urlcolor=black
}

\pgfplotsset{
  grid style = {
    dash pattern = on 0.025mm off 0.95mm on 0.025mm off 0mm, 
    line cap = round,
    black,
    line width = 0.5pt
  },
  tick label style={font=\small},
  label style={font=\small},
  legend style={font=\footnotesize},
}

\pgfplotscreateplotcyclelist{thechairLivingOnTheEdge}{
cyan!60!black,		densely dotted, 	every mark/.append style={fill=cyan!80!black},mark=otimes*\\%
red,                 solid, every mark/.append style={fill=red!80!black},mark=otimes*\\%
cyan!60!black, 		densely dotted,	every mark/.append style={fill=cyan!80!black},mark=diamond*\\%
red,           		solid, every mark/.append style={solid,fill=red!80!black},mark=diamond*\\%
teal,				every mark/.append style={fill=teal!80!black},mark=*\\%
orange,				every mark/.append style={fill=orange!80!black},mark=square*\\%
red!70!white,		mark=star\\%
yellow!60!black,		densely dashed, every mark/.append style={solid,fill=yellow!80!black},mark=square*\\%
black,				every mark/.append style={solid,fill=gray},mark=otimes*\\%
blue,				densely dashed,mark=star,every mark/.append style=solid\\%
red,					densely dashed,every mark/.append style={solid,fill=red!80!black},mark=diamond*\\%
}

\graphicspath{{figures/}}

\bgroup
\makeglossaries
\newacronym{BS}{BS}{base station}
\newacronym{CDN}{CDN}{content delivery network}
\newacronym{CF}{CF}{collaborative filtering}
\newacronym{CN}{CN}{core network}
\newacronym{CRP}{CRP}{{C}hinese restaurant process}
\newacronym{CS}{CS}{central scheduler}
\newacronym{D2D}{D2D}{device-to-device}
\newacronym{GGSN}{GGSN}{Gateway GPRS Support Node}
\newacronym{GPS}{GPS}{global positioning system}
\newacronym{GTP}{GTP}{GPRS Tunneling Protocol}
\newacronym{HetNet}{HetNet}{heterogeneous network}
\newacronym{HTTP}{HTTP}{Hypertext Transfer Protocol}
\newacronym{HD}{HD}{high-definition}
\newacronym{HDFS}{HDFS}{Hadoop Distributed File System}
\newacronym{ICIC}{ICIC}{inter-cell interference coordination}
\newacronym{ICN}{ICN}{information-centric network}
\newacronym{LAC}{LAC}{location area code}
\newacronym{LTE}{LTE}{long term evolution}
\newacronym{MIMO}{MIMO}{multiple-input multiple-output}
\newacronym{massive-MIMO}{massive-MIMO}{massive multiple-input multiple-output}
\newacronym{PDN}{PDN}{packet data network}
\newacronym{PPP}{PPP}{{P}oisson point process}
\newacronym{PHY}{PHY}{physical layer}
\newacronym{RMSE}{RMSE}{root-mean-square error}
\newacronym{RL}{RL}{reinforcement learning}
\newacronym{OTT}{OTT}{over-the-top}
\newacronym{SAC}{SAC}{service area code}
\newacronym{SBS}{SBS}{small base station}
\newacronym{SINR}{SINR}{signal-to-interference-plus-noise ratio}
\newacronym{SCN}{SCN}{small cell network}
\newacronym{SGSN}{SGCN}{Serving GPRS Support Node}
\newacronym{SVD}{SVD}{singular value decomposition}
\newacronym{TEID}{TEID}{tunnel endpoint identifier}
\newacronym{TL}{TL}{transfer learning}
\newacronym{UT}{UT}{user terminal}
\newacronym{URI}{URI}{request-uniform resource identifier}
\newacronym{QoS}{QoS}{quality-of-service}
\newacronym{QoE}{QoE}{quality-of-experience}
\newacronym{RAN}{RAN}{radio access network}

\begin{document}
\title{Big Data Meets Telcos: A Proactive Caching Perspective}
\author{
		\IEEEauthorblockN{Ejder Baştuğ$^{\diamond}$, Mehdi Bennis$^{\star}$, Engin Zeydan$^{\circ}$, Manhal Abdel Kader$^{\diamond}$, Alper Karatepe$^{\circ}$, Ahmet Salih Er$^{\circ}$ and Mérouane Debbah$^{\diamond,\dagger}$				\vspace{0.4cm}} \\
		\IEEEauthorblockA{
				\small
				$^{\diamond}$Large Networks and Systems Group (LANEAS), CentraleSupélec, Université Paris-Saclay, Gif-sur-Yvette, France \\	
				$^{\star}$Centre for Wireless Communications, University of Oulu, Finland \\
				$^{\circ	}$AveaLabs, Istanbul, Turkey \\
				$^{\dagger}$Mathematical and Algorithmic Sciences Lab, Huawei France R\&D, Paris, France \\	
				\small
				ejder.bastug@centralesupelec.fr, 
				bennis@ee.oulu.fi, \\
				\{engin.zeydan, alper.karatepe, ahmetsalih.er\}@avea.com.tr, \\
				merouane.debbah@huawe.com
				manhalak@gmail.com
				\vspace{-0.4cm}
		}
		\thanks{This research has been supported by the ERC Starting Grant 305123 MORE (Advanced Mathematical Tools for Complex Network Engineering), the SHARING project under the Finland grant 128010 and TUBITAK TEYDEB 1509 project grant (numbered  9120067) and the project BESTCOM.}
}
\maketitle
\begin{abstract}
Mobile cellular networks are becoming increasingly complex to manage while classical deployment/optimization techniques and current solutions (i.e., cell densification, acquiring more spectrum, etc.) are cost-ineffective and thus seen as stopgaps. This calls for development of novel approaches that leverage recent advances in storage/memory, context-awareness, edge/cloud computing, and falls into framework of \emph{big data}. However, the big data by itself is yet another complex phenomena to handle and comes with its notorious $4$V: velocity, voracity, volume and variety. In this work, we address these issues in optimization of $5G$ wireless networks via the notion of proactive caching at the base stations. In particular, we investigate the gains of proactive caching in terms of backhaul offloadings and request satisfactions, while tackling the large-amount of available data for content popularity estimation. In order to estimate the content popularity, we first collect users' mobile traffic data from a Turkish telecom operator from several base stations in hours of time interval. Then, an analysis is carried out locally on a big data platform and the gains of proactive caching at the base stations are investigated via numerical simulations. It turns out that several  gains are possible depending on the level of available information and storage size. For instance, with $10\%$ of content ratings and $15.4$ Gbyte of storage size ($87\%$ of total catalog size), proactive caching achieves $100\%$ of request satisfaction and offloads $98\%$ of the backhaul when considering $16$ base stations.
\end{abstract}
%
\section{Introduction}
The unprecedented increase in data traffic demand  driven by mobile video, online social media and \ac{OTT}  applications are compelling mobile operators to look for innovative ways to manage their increasingly complex networks. This explosion of traffic stemming from diverse domain (e.g., healthcare, machine-to-machine communication, connected cars, user-generated content, smart metering, to mention a few) have different characteristics (e.g., structured/non-structured) and is  commonly referred to as \emph{Big Data}  \cite{Lynch2008BigData}. While big data come with "big blessings" there are formidable challenges in dealing with large-scale data sets due to the sheer volume and dimensionality of the data.  A fundamental challenge of big data analytics is to shift through  large volumes of  data  in order to discover hidden patterns for actionable decision making. Indeed, the era of collecting and storing  data in remote standalone servers where  decision making is  done  offline has dawned. Rather, telecom operators are  exploring decentralized  and flexible network architectures whereby predictive resource management  play a crucial role leveraging recent advances in storage/memory, context-awareness and edge/cloud computing \cite{Bastug2014LivingOnTheEdge, Bonomi2012Fog, Luan2015Fog}.  In the realm of wireless, big data brings to network planning a variety of new information sets that can be inter-connected to achieve a better understanding of users and networks (e.g.,  location, user velocity, social geodata, etc.). Moreover, public data from social networks such as Twitter and Facebook provides additional side information about the life of the network, which can be further exploited. The associated benefits are a higher accuracy of user location information or the ability to easily identify and predict user clustering, for example for special events. Undoubtedly, the huge potential associated with big data has sparked a flurry of research interest from industry, government and academics (see \cite{Hu2014TowardScalable} for a recent survey), and will continue to do so in the coming years.

At the same time, mobile cellular networks are evolving towards the next generation of  $5$G wireless communication, in which  ultra-dense networks, millimetre wave communications, \ac{massive-MIMO}, edge caching, device-to-device communications  play a pivotal role (see \cite{Andrews2014Will} and references therein). Unlike the base station-centric architecture  paradigm assuming  \emph{dumb} terminals and in which network optimization is carried out  in a \emph{reactive} way, $5$G networks will be truly disruptive in terms of being user-centric, context-aware and proactive/anticipatory in nature. While continued evolution in spectral efficiency is  expected, the maturity of air interfaces of current systems (LTE-Advanced) mean that no major improvements of spectral efficiency can be anticipated. Additional measures like the brute force expansion of wireless infrastructure (number of cells) and the licensing of more spectrum are prohibitively expensive. Thus, innovative solutions are called upon.

In this work, based on the motivations and issues above, we are intent to propose a proactive caching architecture for optimization of $5$G wireless networks where we exploit large amount of available data with the help of big data analytics and machine learning tools. In other words, we investigate the gains of proactive caching both in terms of backhaul offloadings and request satisfactions, where machine learning tools are used to model and  predict the spatio-temporal user behaviour for proactive cache decision. By caching strategic contents at the edge of network, namely at the base stations, network resources are utilized more efficiently and users' experience is further improved. However, the estimation of content popularity tied with spatio-temporal behaviour of users is a very complex problem due to the high dimensional aspects of data, data sparsity and lack of measurements. In this regard, we present a platform to parallelize the computation and execution of the content prediction algorithms for cache decision at the base stations. As a real-world case study, a large amount of data collected from a Turkish telecom operator, one of the largest mobile operator in Turkey with $16.2$ million of active subscribers, is examined for various caching scenarios. Particularly, the traces of mobile users' activities are collected from several base stations in hours of time interval and are analysed inside the network under the privacy concerns and regulations. The analysis is carried out on a big data platform and caching at the base stations has been investigated for further improvements of users' experience and backhaul offloadings.
\subsection{Prior Work and Our Contribution}
The use of big data in mobile computing research has been investigated recently such as in \cite{Laurila2012BigDataMobile}. The idea of caching at the edge of wireless network has also been studied in various works \cite{Bastug2013Proactive, Golrezaei2013FemtocachingD2D, Bastug2014CacheEnabledExtended, Hamidouche2014ManyToMany, Poularakis2014Approximation, Zhou2015Multicast, Nagaraja2015Caching}, including proactive caching for $5$G wireless networks \cite{Bastug2014LivingOnTheEdge}. In detail, a proactive caching procedure using perfect knowledge of content popularity is studied in \cite{Bastug2013Proactive}. A caching architecture (namely FemtoCaching) relying on cache-enabled user devices and small base stations is introduced in \cite{Golrezaei2013FemtocachingD2D}. The caching problem as a many-to-many matching game is formulated in \cite{Hamidouche2014ManyToMany} and caching gains are characterized numerically. Deployment aspects of cache-enabled base stations via stochastic geometry tools is investigated in \cite{Bastug2014CacheEnabledExtended} where the outage probability is derived as a function of \ac{SINR}, base station density and storage size. For optimal cache allocations, an approximation framework based on a well-known facility location problem is given in \cite{Poularakis2014Approximation}. The impact of unknown content popularity on cache decision is characterized in \cite{Nagaraja2015Caching}. The advantage of multicast transmission together with caching at the base station is investigated in \cite{Zhou2015Multicast}. We refer our readers to \cite{Bastug2015ThinkBefore} for a recent survey and more comprehensive details.

Compared to the works mentioned above, our main contribution in this work is to make tighter connections of big data phenomena with caching in $5G$ wireless networks, by proposing a proactive caching  architecture where statistical machine learning tools are exploited for content popularity estimation. Combined with a large-scale real-world case study, this is perhaps the first attempt on this direction and highlights a huge potential of big data for $5$G wireless networks. 

The rest of paper is organized as follows. Our network model for proactive caching is detailed in Section \ref{sec:networkmodel}. A practical case study of content popularity estimation on a big data platform is presented in Section \ref{sec:bigdata}, including a characterization of users' traffic pattern. Subsequently, numerical results for cache-enabled base stations and relevant discussions are carried out in Section \ref{sec:numerical}. We finally conclude in Section \ref{sec:conclusions} and draw our future directions in the same section.
\section{Network Model}
\label{sec:networkmodel}
\begin{figure*}[ht!]
	\centering
	\includegraphics[width=0.95\linewidth]{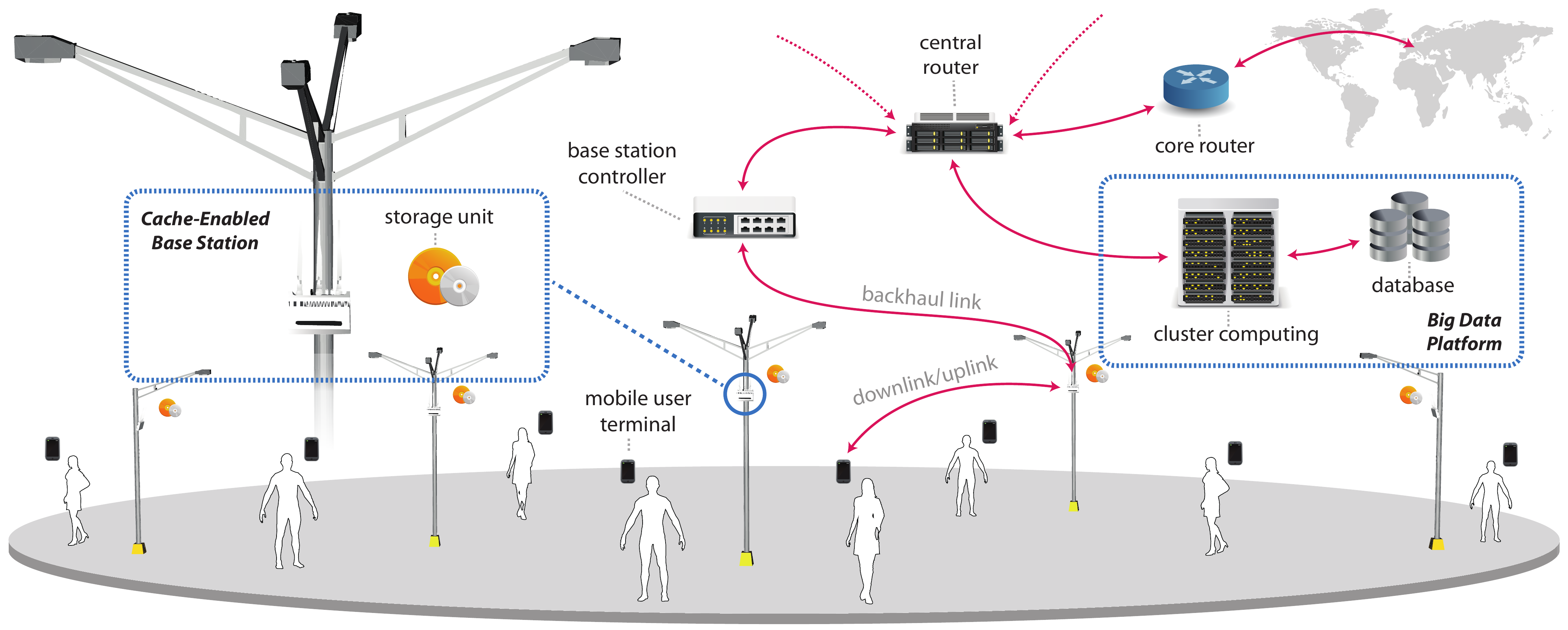}
	\caption{An illustration of the network model. A \emph{big data platform} is in charge of tracking/predicting users' demand, whereas \emph{cache-enabled base stations} store the strategic contents predicted on the big data platform.}
	\label{fig:architecture}
\end{figure*} 
Suppose a network deployment of $M$ \glspl{SBS} from the set $\mathcal{M} = \lbrace 1, \ldots, M \rbrace$ and $N$ \glspl{UT} from the set $\mathcal{N}= \lbrace 1, \ldots, N\rbrace$. Each \ac{SBS} $m$ has access to the broadband Internet connection via a wired backhaul link with capacity $C_m$ Mbyte/s, and is able to provide this broadband service to its users via a wireless link with total capacity of  $C'_m$ Mbyte/s. Due to the motivation that the backhaul capacity is generally limited in densely deployed \glspl{SBS} scenarios \cite{Andrews2014Will}, we further consider that $C_m < C'_m$. Also, assume that each user $n \in \mathcal{N}$ is connected to only one \ac{SBS} and is served via unicast sessions\footnote{The unicast service model can also be extended to the multicast case. See \cite{Poularakis2014Multicast, Zhou2015Multicast} for studies in this direction.}. In particular, we assume that \glspl{UT} request contents (i.e., videos, files, news, etc.) from a library $\mathcal{F} = \lbrace 1, \ldots, F\rbrace$, where each content $f$ in this library has a size of $L(f)$ Mbyte and bitrate requirement of $B(f)$  Mbyte/s, with
\begin{eqnarray}
	L_{\text{min}} &= \underset{f \in \mathcal{F}}{\text{min}} \{ L(f) \} > 0  \\
	L_{\text{max}} &= \underset{f \in \mathcal{F}}{\text{max}} \{ L(f) \} < \infty
\end{eqnarray}
and
\begin{eqnarray}
	B_{\text{min}} &= \underset{f \in \mathcal{F}}{\text{min}} \{ B(f) \} > 0 \\
	B_{\text{max}} &= \underset{f \in \mathcal{F}}{\text{max}} \{ B(f) \} < \infty.
\end{eqnarray}
The users' content requests in fact follow a Zipf-like distribution $P_{\mathcal{F}}(f), \forall f \in \mathcal{F}$ given as \cite{Breslau1999WebZipf}:
\begin{equation}\label{eq:zipf}
	P_{\mathcal{F}}(f) = \frac{\Omega}{f^{\alpha}}
\end{equation}
where 
\begin{equation}
	\Omega = \Big(\sum_{i=1}^{F}{\frac{1}{i^{\alpha}}}\Big)^{-1} \nonumber.
\end{equation}
The parameter $\alpha$ in \eqref{eq:zipf} describes the steepness of the distribution. This kind of power laws is used to characterize many real-world phenomena, such as the distribution of files in the web-proxies \cite{Breslau1999WebZipf} and the traffic dynamics of cellular devices \cite{Shafiq2011Characterizing}. Higher values of $\alpha$ corresponds to a steeper distribution, meaning that a small subset of contents are highly popular than the rest of the catalog (namely users have very similar interests). On the other hand, the lower values describe a more uniform behaviour with almost equal popularity of contents (namely users have more distinct interests). The parameter $\alpha$ can take different values depending on users' behaviour and \glspl{SBS} deployment strategies (i.e., home, enterprise, urban and rural environments), and its practical value in our experimental setup will be given in the subsequent sections.

Given such a global content popularity in the decreasing ordered case, the content popularity matrix of the $m$-th \ac{SBS} at time $t$ is specifically described by ${\mathbf P}^{m}(t) \in \mathbb{R}^{N \times F}$ where each entry $P^m_{n,f}(t)$ corresponds to the probability that the $n$-th user requests the $f$-th content. In fact, the matrix ${\bf P}^m (t)$ is the local content popularity distribution observed at the base station $m$ at time $t$, whereas the Zipf distribution $P_{\mathcal{F}}(f), \forall f \in \mathcal{F}$ is used to characterize the global content popularity distribution of all contents  in (decreasing) sorted order.

In this scenario, we consider that each \ac{SBS} has a finite storage capacity of $S_m$ and proactively caches selected contents from the library $\mathcal{F}$ during peak-off hours. By doing so, the bottlenecks caused by the limited-backhaul are avoided during the delivery of users' content requests in peak hours. The amount of satisfied requests and backhaul load are of paramount importance and are defined as follows. Suppose that $D$ number of contents are requested during the duration of $T$ seconds, and are represented by the set $\mathcal{D} = \{1, ..., D\}$. Assume that the delivery of content is started immediately when the request $d \in \mathcal{D}$ arrives to the \ac{SBS}. Then, the request $d$ is called \emph{satisfied} if the rate of content delivery is equal or higher than the bitrate of the content in the end of service, such as:
\begin{equation}\label{eq:satisfied}
	\frac{L(f_d)}{\tau'(f_d)  - \tau(f_d)} \ge B(f_d)
\end{equation}
where $f_d$ describes the requested content, $L(f_d)$ and $B(f_d)$ are the size and bitrate of the content, $\tau(f_d)$ is the arrival time of the content request and $\tau'(f_d)$ the end time delivery.\footnote{One can also consider/exploit future information (i.e., start time of requests, end time of content delivery) in the context of proactive resource allocation (see \cite{Tadrous2014Proactive} for instance).} Defining the condition in \eqref{eq:satisfied} stems from the fact that, if the delivery rate is not equal nor higher than the bitrate of the requested content, the interruption during the playback (or download) occurs thus users would have less \ac{QoE}\footnote{In practice, a video content has typically a bitrate requirement ranging from $1.5$ to $68$ Mbit/s \cite{Google2015Bitrates}.}. Therefore, the situations where this condition holds are more desirable for better \ac{QoE}. In \eqref{eq:satisfied}, note also that the end time of delivery for request $d$, denoted by $\tau'(d)$, highly depends on the load of the system, capacities of the backhaul and wireless links as well as availability of contents at the base stations. Given this definition of satisfied requests and related explanations, the users' average request \emph{satisfaction ratio} is then defined for the set of all requests, that is:
\begin{equation}
	\eta(\mathcal{D}) = \frac{1}{D}\sum_{d \in \mathcal{D}}{\mathds{1} \left\lbrace \frac{L(f_d)}{\tau'(f_d)  - \tau(f_d)} \ge B(f_d) \right\rbrace}
\end{equation}
where $\mathds{1}\left\lbrace ... \right\rbrace$ is the indicator function which takes $1$ if the statement holds and $0$ otherwise. Now, denoting $R_d(t)$ Mbyte/s as the instantaneous rate of backhaul for the request $d$ at time $t$, with  $R_d(t) \leq C_m$, $\forall m \in \mathcal{M}$, the average \emph{backhaul load} is then expressed as:
\begin{equation}
	\label{eq:bload}
	\rho(\mathcal{D}) = \frac{1}{D}\sum_{d \in \mathcal{D}}{\frac{1}{L(f_d)}\sum_{t=\tau(f_d)}^{\tau'(f_d)}{R_d(t)}}.
\end{equation}
Here, the outer sum is over the set of all requests whereas the inner sum gives the total amount of information passed over the backhul for request $d$ which is at most equal to the length of requested file $L(f_d)$. The instantaneous rate of backhul for request $d$, denoted by $R_d(t)$, heavily depends on the load of the system, capacity of the backhaul link and cached contents at the base stations. 

In fact, by pre-fetching the contents at the \glspl{SBS}, the access delays to the contents are minimized especially during the peak hours, thus yielding higher satisfaction ratio and less backhaul load. To elaborate this, now consider the cache decision matrix of \glspl{SBS} as ${\bf X}(t) \in \{0, 1\}^{M\times F}$, where the entry $x_{m,f}(t)$ takes $1$ if the $f$-th content is cached at the $m$-th \ac{SBS} at time $t$, and $0$ otherwise. Then, the backhaul offloading problem under a specific request satisfaction constraint is formally given as follows:
\begin{align}
\label{eq:problem1}
&	\underset{{\bf X}(t), {\bf P}^m(t)}{\text{minimize}}	&	& \rho(\mathcal{D}) 			     \\
&	\text{subject to}	& & L_{\text{min}} \leq L(f_d) \leq L_{\text{max}},	\hspace{2.2cm}   \forall d \in \mathcal{D}, \tag{\ref{eq:problem1}a} \\
&						& & B_{\text{min}} \leq B(f_d) \leq B_{\text{max}}, 	\hspace{2.15cm}   \forall d \in \mathcal{D},\tag{\ref{eq:problem1}b} \\
&						& & R_d(t) \leq C_m, \hspace{1.2cm} \forall t, \forall d \in \mathcal{D}, \forall m \in \mathcal{M},\tag{\ref{eq:problem1}c} \\
&						& & R'_d(t) \leq C'_m, \hspace{1.2cm} \forall t, \forall d \in \mathcal{D}, \forall m \in \mathcal{M}, \tag{\ref{eq:problem1}d} \\
&						& & \sum_{f \in \mathcal{F}} L(f)x_{m,f}(t) \leq S_m, 	\hspace{0.4cm}  \forall t,  \forall m \in \mathcal{M},\tag{\ref{eq:problem1}e} \\
&						& & \sum_{n \in \mathcal{N}}\sum_{f \in \mathcal{F}}{P^m_{n,f}(t)} = 1, 		   \hspace{0.45cm} \forall t, \forall m \in \mathcal{M}, \tag{\ref{eq:problem1}f} \\
&						& & x_{m,f}(t) \in \{0, 1\},  \hspace{0.2cm} 		 \forall t, \forall f \in \mathcal{F}, \forall m \in \mathcal{M}, \tag{\ref{eq:problem1}g} \\
&						& & \eta_{\text{min}} \leq \eta(\mathcal{D}), 		   \tag{\ref{eq:problem1}h}
\end{align}
where $R'_d(t)$ Mbyte/s describes the instantaneous rate of wireless link for request $d$ and $\eta_{\text{min}}$ represents  the minimum target satisfaction ratio. In particular, the constraints (\ref{eq:problem1}a) and (\ref{eq:problem1}b) are to bound the length and bitrate of contents in the catalog for feasible solution, the constraints (\ref{eq:problem1}c) and (\ref{eq:problem1}d) are the backhaul and wireless link  capacity constraints, (\ref{eq:problem1}e) holds for storage capacity for caching, (\ref{eq:problem1}f) is to ensure the content popularity matrix as a probability measure, (\ref{eq:problem1}g) denotes the binary decision variables of caching, and finally the expression in (\ref{eq:problem1}h) is the satisfaction ratio constraint for \ac{QoE}.

In order to tackle this problem, the cache decision matrix ${\bf X}(t)$ and the content popularity matrix estimation ${\bf P}^m(t)$ have to be optimized jointly. However, solving the problem (\ref{eq:problem1}) is very challenging as:
\begin{itemize}
\item[i)]  the storage capacity of \glspl{SBS}, the backhaul and wireless link capacities are limited.
\item[ii)] the catalog size and number of users with unknown ratings\footnote{The term "rating" refers to the empirical value of content popularity/probability and is interchangeable throughout the paper.} are very large in practice. 
\item[iii)] the optimal uncoded\footnote{In the information theoretical sense, the caching decision can be categorized into "coding" and "uncoded" groups (see  \cite{Maddah2014Fundamental} for example).} cache decision for a given demand is non-tractable \cite{Bastug2013Proactive, Poularakis2014Approximation, Golrezaei2013FemtocachingD2D}.
\item[iv)] the \glspl{SBS} have to track, learn and estimate the sparse content popularity/rating matrix \glspl{SBS}  ${\mathbf P}^m(t)$ while making the cache decision.
\end{itemize}
In order to overcome these issues, we restrict ourselves to the fact that cache decision is made during peak-off hours, thus ${\bf X}(t)$ remains static during the content delivery in peak hours and is represented by ${\bf X}$. Additionally, the content popularity matrix is stationary during $T$ time slots and identical among the base stations, thus ${\mathbf P}^m(t)$ is represented by ${\mathbf P}$. 

After these considerations, we now suppose that the problem can be decomposed into two parts in which the content popularity matrix ${\bf P}$ is first estimated, then is used in the caching decision ${\bf X}$ accordingly. In fact, if sufficient amount of users' ratings are available at the \glspl{SBS}, we can construct a $k$-rank approximate popularity matrix ${\mathbf P} \approx  {\mathbf N}^T{\mathbf F}$, by jointly learning the factor matrices ${\mathbf N} \in \mathbb{R}^{k \times N}$ and ${\mathbf F} \in \mathbb{R}^{k \times F}$ that minimizes the following cost function:
\begin{align}
	\label{eq:classical}
	\underset{{\mathbf P}}{\text{minimize}}	& \sum_{P_{ij} \in {\mathcal P}} \Big({\mathbf n}_i^T{\mathbf f}_j-P_{ij}\Big)^2 + \mu\Big(||{\mathbf N}||^2_F+||{\mathbf F}||^2_F\Big) 		
\end{align}
where the summation is done over the corresponding user/content rating pairs $P_{ij}$ in the training set ${\mathcal P}$. The vectors ${\mathbf n}_i$ and ${\mathbf f}_j$ here describe the $i$-th and $j$-th columns of ${\mathbf N}$ and ${\mathbf F}$ matrices respectively, and $||.||^2_F$ represents the Frobenius norm. The parameter $\mu$ is used to provide a balance between the regularization and fitting the training data. Therein, high correspondence between the user factor matrix ${\mathbf N}$ and content factor matrix ${\mathbf F}$ leads to a better estimate of ${\mathbf P}$. In fact, the problem \eqref{eq:classical} is a regularized least square problem where the matrix factorization is embedded in the formulation. Despite various approaches, the matrix factorization methods are commonly used to solve this kind of problems and has many applications such as in recommendation systems (i.e., Netflix video recommendation).  In our case detailed in the following sections, we have used regularized sparse \ac{SVD} to solve the problem algorithmically which exploits the least square nature of the problem. The overview of these approaches, sometimes called \ac{CF} tools, can be found in \cite{Koren2009Matrix, Lee2012CF}. When the estimation of content popularity matrix ${\bf P}$ is obtained, the caching decision ${\bf X}$ can be made in this scenario accordingly. 

In practice, the estimation of ${\mathbf P}$ in \eqref{eq:classical} can be done by collecting/analysing large amount of available data on a \emph{big-data platform} of the network operator, and strategic/popular contents from this estimation can be stored at the \emph{cache-enabled base stations} whose cache decisions are represented by ${\bf X}$. By doing this, the backhaul offloading problem in \eqref{eq:problem1} is minimized and higher satisfactions are achieved. Our network model including such an infrastructure is illustrated in Fig. \ref{fig:architecture}. In the following, as a case study, we detail our big data platform and present users' traffic characteristics by analysing large amount of data on this platform. The processed data will be used to estimate the content popularity matrix ${\mathbf P}$ which is essentially required for the cache decision ${\bf X}$ and will be detailed in the upcoming sections.
\section{Big Data Platform}
\label{sec:bigdata}
The big data platform used in this work runs in the operator's core network. As mentioned before, the purpose of this platform is to store users' traffic data and extract useful information which are going to be used for content popularity estimation.
In a nutshell, the operator's network consists of several districts with more than $10$ regional core areas  throughout Turkey. The average total traffic over all regional areas consists of approximately over $15$ billion packets in uplink direction and over $20$ billion packets in the downlink direction daily. This corresponds to approximately over $80$~TByte of total data flowing in uplink and downlink daily in a mobile operator’s core network. The data usage behaviour results in exponential increase in data traffic of a mobile operator. For example, in $2012$, the approximate total data traffic was over $7$ TByte in both uplink and downlink daily traffic. 

The streaming traces which will be detailed in the sequel, are obtained from one of the operator's core network region, includes the mobile traffic from many base stations, and are captured by a server on a high speed link of $200$ Mbit/sec at peak hours. In order to capture Internet traffic data by the server in this platform, a procedure is initialized by mirroring real-world Gn interface data.\footnote{Gn is an interface between \ac{SGSN} and \ac{GGSN}. Network packets sent from a user terminal to the \ac{PDN}, e.g. internet, pass through \ac{SGSN} and \ac{GGSN} where \ac{GTP} constitutes the main protocol in network packets flowing through Gn interface.} After mirroring stage of Gn interface, network traffic is transferred into the server on the platform. For our analysis, we have collected traffic of approximately $7$ hours starting from $12$ pm to $7$ pm on Saturday $21$'st of March $2015$. This traffic is processed on the big data platform which is essentially based on Hadoop.
%
\subsection{Hadoop platform}
Among the available platforms, Hadoop stands out as the most notable one as it is an open source solution~\cite{Hadoop}. It is made up of a storage module, namely \ac{HDFS} and a computation module, namely MapReduce. Whereas \ac{HDFS} can have centralized or distributed implementations, MapReduce inherently has a distributed structure that enables it to execute jobs in parallel on multiple nodes.

As stated in previous subsection, the accuracy and precision of the proposed mechanism was tested in operator's network. A data processing platform was implemented through using Cloudera's Distribution Including Apache Hadoop (CDH4)~\cite{Cloudera} version on four nodes including one cluster name node, with computations powers corresponding to each node with INTEL Xeon CPU E5-2670 running @2.6 GHz, $32$ Core CPU, $132$ GByte RAM, $20$ TByte hard disk. This platform is used to extract the useful information from raw data which is described as follows.
%
\subsection{Data extraction process}
First, the raw data is parsed using Wireshark command line utility \emph{tshark}~\cite{Wireshark} in order to extract the relevant fields of CELL-ID (or \ac{SAC} in our case, in order to uniquely identify a \emph{service area} within a \emph{location area}\footnote{The service area identified by \ac{SAC} is an area of one or more base stations, and belongs to a location area which is uniquely identified by \ac{LAC}. Typically, tens or even hundreds of base stations operates in a given location area.}), \ac{LAC}, \ac{HTTP} \ac{URI},  \ac{TEID}\footnote{A \ac{TEID} uniquely identifies a tunnel endpoint on the receiving end of the GTP tunnel. A local \ac{TEID} value is assigned at the receiving end of a \ac{GTP} tunnel in order to send messages through the tunnel.}  and \ac{TEID}-DATA for data and control plane packets respectively, and FRAME TIME indicating arrival time of packets.  The \ac{HTTP} Request-URI is a Uniform Resource Identifier that identifies the resource upon which to apply the request. The \emph{control} packets contain the information elements that carry the information required for future data packets. It contains cell identification ID (CELL-ID), \ac{LAC} and \ac{TEID}-DATA fields. The \emph{data} packets contain \ac{HTTP}-\ac{URI} and \ac{TEID} fields. 

In the next step, after obtaining those relevant fields from both control and data packets, the extracted data is transferred into \ac{HDFS} for further analysis. In \ac{HDFS}, there can be done many data analytics performed over the collected data using  Hive Query language (QL)~\cite{Hive}.  For example, in order to calculate the \ac{HTTP} Request-\glspl{URI} at specific location, the \ac{HTTP}-\ac{URI} can be joined with CELL-ID-\ac{LAC} fields over the same \ac{TEID} and \ac{TEID}-DATA fields for data and control packets respectively. In our analysis, due to the limitations on observable number of rows of \ac{HTTP}-\ac{URI} fields with a corresponding CELL-ID-\ac{LAC} fields after mapping, we have proceeded with \ac{HTTP} Request-\glspl{URI} and \ac{TEID} mappings.

\begin{figure}[ht!]
	\centering
	\includegraphics[width=0.7\linewidth]{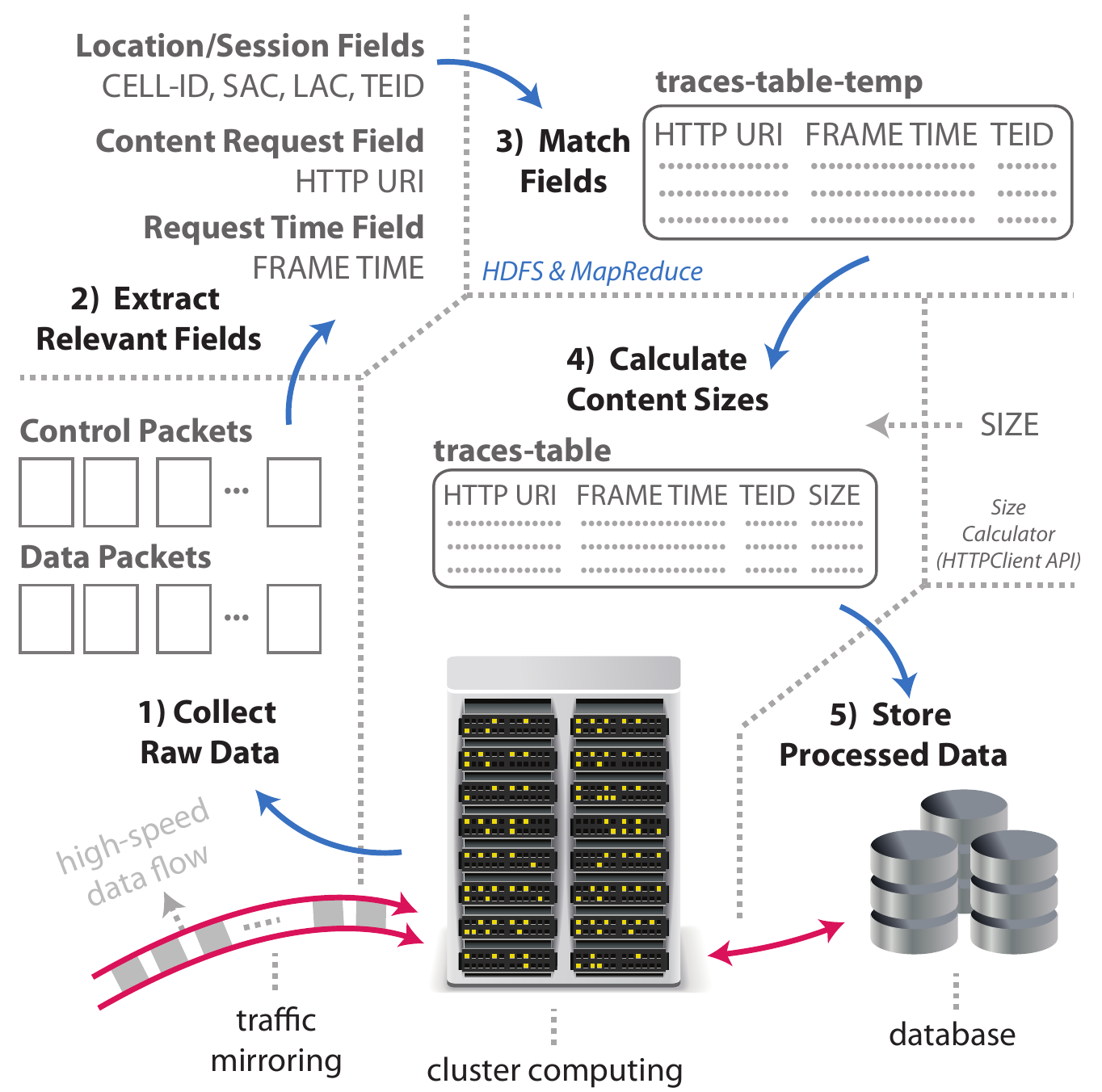}
	\caption{An overview of the data extraction process on the big data platform.}
	\label{fig:extraction}
\end{figure}

From \ac{HDFS}, a temporary table named \textit{traces-table-temp} is constructed using Hive QL. The \textit{traces-table-temp} has  \ac{HTTP} Request-\ac{URI}, FRAME TIME and \ac{TEID} fields. After constructing this table, the sizes of each \ac{HTTP} Request-\ac{URI} request is calculated using a separate \ac{URI}-\emph{size calculator} program that uses HTTPClient API~\cite{HTTP} in order to obtain the final table  called \textit{traces-table} with fields of SIZE, \ac{HTTP} Request-\glspl{URI}, FRAME TIME and \ac{TEID}. This table has approximately over $420.000$ of $4$ millions \ac{HTTP} Request-\ac{URI}'s with SIZE field returned as not zero or null due to unavailability of \ac{HTTP} response for some requests. Note that in a given session with a specific \ac{TEID}, there can be multiple \ac{HTTP} Request-\glspl{URI}. Each \ac{TEID} belongs to specific user. Each user can also have multiple \glspl{TEID} with multiple \ac{HTTP} Request-\glspl{URI}. The steps of data extraction process on the platform is summarized in Fig. \ref{fig:extraction}. Note that the data extraction process is specific to our scenario for proactive caching. However, similar studies in terms of usage of big data platform and exploitation of big data analytics for telecom operators can be found in \cite{Dong2011Teledata, Jeong2012Anomaly, Magnusson2012Subscriber, Indyk2012Mapreduce, Celebi2013BigData, Karatepe2014Anomaly}.
%
\subsection{Traffic Characteristics}
\begin{figure*}[!ht]
\centering
\begin{subfigure}{.45\textwidth}
\centering
\begin{tikzpicture}
	\begin{axis}[
		width=\textwidth,
 		grid = major,
 		legend cell align=left,
 		xmode=log,
 		ymode=log,
 		legend style ={legend pos=north east},
 		xlabel={Rank},
 		ylabel={Nr. of occurence}]

 		\addplot+[cyan!60!black, only marks, mark=o, mark size=1.8, thick]
 				  table [col sep=comma] {\string"results-globalpop-c1trace.csv"};
 		\addlegendentry{Collected traces};

		\addplot+[red!80!black, dashed, very thick, mark=none] 
                                 table [col sep=comma] {\string"results-globalpop-c2fit1.36.csv"};
		\addlegendentry{Zipf fit, 	$\alpha = 1.36$};
		 		  		
	\end{axis}
\end{tikzpicture}
\caption{Global content popularity distribution.}
\label{fig:globalpop}
\end{subfigure}
\hspace{0.8cm}
\begin{subfigure}{.45\textwidth}
\centering
\begin{tikzpicture}
	\begin{axis}[
		width=\textwidth,
 		grid = major,
 		legend cell align=left,
 		xmode=log, 		
 		mark repeat={10},
 		legend style ={legend pos=north east},
 		xlabel={ Rank},
 		ylabel={Cumulative Size [GByte]}]

 		\addplot+[blue!60!black, mark=o, mark size=1.8, thick]
 				  table [col sep=comma] {\string"results-globalpop-cumsize.csv"};
		 		  		
	\end{axis}
\end{tikzpicture}
\caption{Cumulative size distribution.}
\label{fig:globalpop-cumsize}
\end{subfigure}
%
\caption{Behaviour of content popularity distribution.}
\end{figure*}
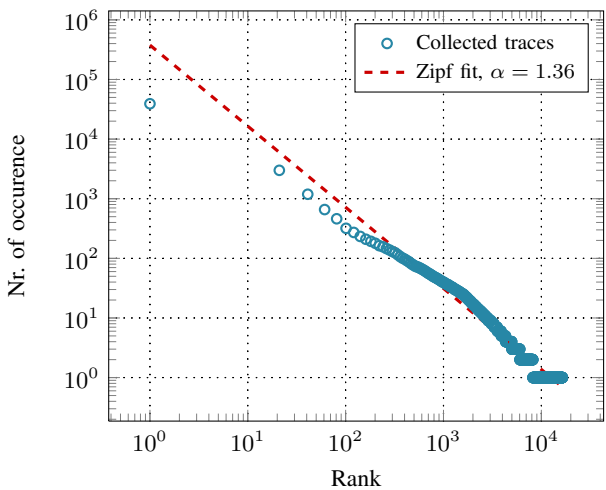
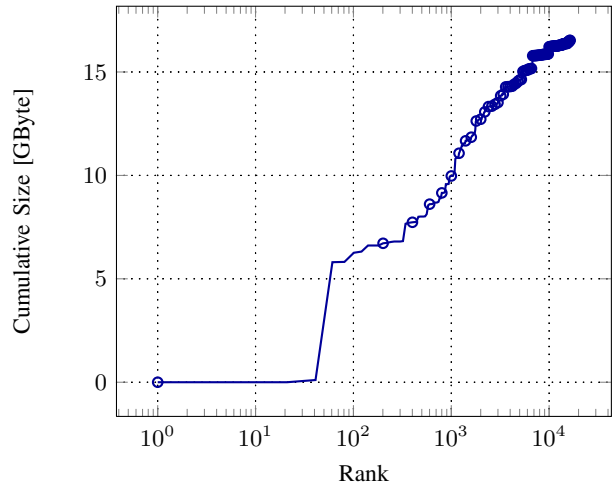
Based on information available in \textit{traces-table}, the global content popularity distribution (namely \ac{HTTP}-\ac{URI} popularity distribution) in a decreasing ranked order is plotted in Fig. \ref{fig:globalpop}. According to this available experimental data, we observe that the popularity behaviour of contents follows a Zipf law with steepness parameter $\alpha = 1.36$.\footnote{The value of steepness parameter $\alpha$ can change depending on the scenario. For instance,  the steepness parameter of content popularities in YouTube catalog varies from $1.5$ to $2.5$ \cite{Cha2007Tube, Rossi2012Sizing}.} Therein, the Zipf curve is calculated in the  least square sense from the collected traces and the parameter $\alpha$ is then found by evaluating the slope of the curve. On the other hand, cumulative size of ranked contents is given in Fig. \ref{fig:globalpop-cumsize}. The cumulative size up to $41$-th most-popular contents has $0.1$ GByte of size,  whereas a dramatical increase appears afterwards. This basically shows that most of the requested contents in our traces has low content sizes and contents with larger sizes are relatively less requested.

We would like to note that a detailed characterization of the traffic for caching is left for future work. Indeed, characterization of the traffic in web proxies which are placed in the intermediate level of network \cite{Breslau1999WebZipf}, a specific video content catalog in a campus network \cite{Zink2009Characteristics}, mobile traffic of users in Mexico \cite{Oliveira2014Measurement} can be found in the literature. Compared to these works, we focus on the characterization traffic of mobile users collected from base stations in a large regional area and exploit this information for proactive caching (i.e., content popularity distribution, cumulative size distribution). Based on information available in \textit{traces-table}, we in the following simulate a scenario of cache-enabled base stations.
\section{Numerical Results and Discussions}
\label{sec:numerical}
The list of parameters for numerical setup is given in Table \ref{tab:setup_params1}. For ease of analysis, the storage, backhaul, and wireless link capacities of small cells are assumed to be identical within each other. 
%
\begin{table}[ht]
\centering
\scriptsize
\caption{List of simulation parameters.\label{tab:setup_params1}}
\begin{tabular}{|c|l|c|}
\hline
\textbf{Parameter}			  &  \textbf{Description} 	& 	\textbf{Value}\\
\hline
$T$    		& Time slots							& 	$ 6$ hours $47$ minutes \\
\hline
$D$ 			& Number of requests 				& 	$422529$ 	\\
\hline
$F$				& Number of contents 			&  $16419$	 \\
\hline
$M$   			& Number of small cells 				& 	$16$ \\
\hline
$L_{\mathrm{min}}$  			& Min. size of a content & 	$1$	Byte \\
$L_{\mathrm{max}}$  			& Max. size of a content & 	$6.024$	GByte \\
\hline
$B(f)$ 			& Bitrate of content $f$				&	$4$ Mbyte/s	\\
\hline
$\sum_{m}{C_m}$ 			& Total backhaul link capacity		&	$3.8$	Mbyte/s	\\
\hline
$\sum_{m}\sum_{n}C'_{m}$  	& Total wireless link capacity 		&	$120$	 Mbyte/s\\
\hline
\end{tabular}
\end{table}

In the simulations, all of $D$ number of requests are taken from the processed data (namely \emph{traces-table}), spanning over a time duration of $6$ hours $47$ minutes. The arrival times of each request (FRAME TIME), requested content (\ac{HTTP}-\ac{URI}) and content size (SIZE) are taken from the same table. Then, these requests are associated to $M$ base stations pseudo-randomly. In order to solve the backhaul offloading problem in \eqref{eq:problem1}, the content popularity matrix ${\bf P}$ and caching strategy ${\bf X}$ are evaluated separately. In particular, the following two methods are used for constructing the content popularity matrix ${\bf P}$: 
\begin{itemize}
	\item \emph{Ground Truth}: The content popularity matrix ${\bf P}$ is constructed from all available information in \emph{traces-table} instead of solving the problem in \eqref{eq:classical}. Note that the rows of ${\bf P}$ represent base stations and columns are contents. The rating density of this matrix is $6.42\%$. 
	\item \emph{Collaborative Filtering}: For the estimation of content popularity matrix ${\bf P}$, the problem in \eqref{eq:classical} is attempted by first choosing $10\%$ of ratings  from \emph{traces-table} uniformly at random. Then, these ratings are used in the training stage of the algorithm and missing entries/ratings of ${\bf P}$ are estimated. Particularly, the regularized \ac{SVD} from the \ac{CF} methods \cite{Paterek2007Improving, Lee2012CF} is used in the algorithmic part.
\end{itemize}
After constructing the content popularity matrix ${\bf P}$ based on these above methods, the cache decision (modelled by the matrix ${\bf X}$) is made by storing the most-popular contents greedily at the \glspl{SBS} until no storage space remains (see \cite{Bastug2013Proactive} for the details). Having these contents cached proactively at the \glspl{SBS} at $t=0$, the requests are then served until all of the contents are delivered. The performance metrics request satisfaction and backhaul load are calculated accordingly.

\begin{figure*}[!ht]
\centering
\begin{subfigure}{.45\textwidth}
\centering
\begin{tikzpicture}
	\begin{axis}[
		width=\textwidth,
 		grid = major,
 		legend cell align=left,
 		xmin=0,xmax=100,	
 		legend style ={legend pos=south east},
 		xlabel={Storage Size (\%)},
 		ylabel={Request Satisfaction (\%)}]

 		\addplot+[smooth, blue!60!black, mark=o, mark size=1.8, thick]
 				  table [col sep=comma] {\string"results-satisfaction-c1ground.csv"};
 		\addlegendentry{Ground Truth};

		\addplot+[red!80!black, dashed, very thick, mark=none] 
                                 table [col sep=comma] {\string"results-satisfaction-c2cf.csv"};
		\addlegendentry{Collaborative Filtering};
		 		  		
	\end{axis}
\end{tikzpicture}
\caption{Evolution of satisfaction with respect to the storage size.}
\label{fig:satisfaction}
\end{subfigure}
\hspace{0.8cm}
\begin{subfigure}{.45\textwidth}
\centering
\begin{tikzpicture}
	\begin{axis}[
		width=\textwidth,
 		grid = major,
 		legend cell align=left,
 		xmin=0,xmax=100,	
 		legend style ={legend pos=north east},
 		xlabel={Storage Size (\%)},
 		ylabel={Backhaul Load (\%)}]

 		\addplot+[blue!60!black, mark=o, mark size=1.8, thick]
 				  table [col sep=comma] {\string"results-backhaul-c1ground.csv"};
 		\addlegendentry{Ground Truth};

		\addplot+[red!80!black, dashed, very thick, mark=none] 
                                 table [col sep=comma] {\string"results-backhaul-c2cf.csv"};
		\addlegendentry{Collaborative Filtering};
		 		  		
	\end{axis}
\end{tikzpicture}
\caption{Evolution of backhaul usage with respect to the storage size.}
\label{fig:offloading}
\end{subfigure}
%
\caption{Simulation results of proactive caching at the base stations.}
\label{fig:satback}
\end{figure*}
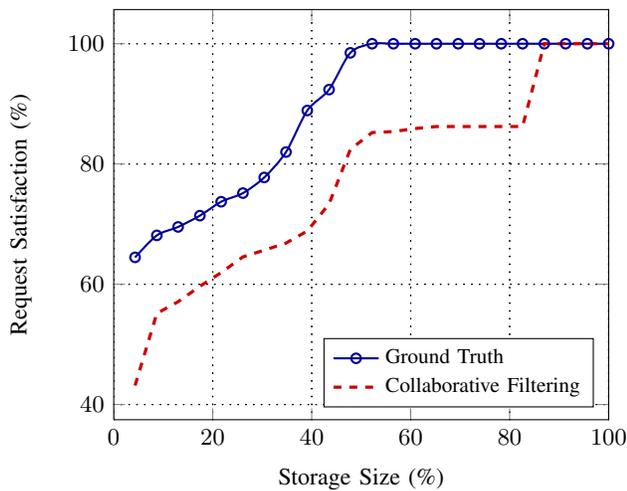
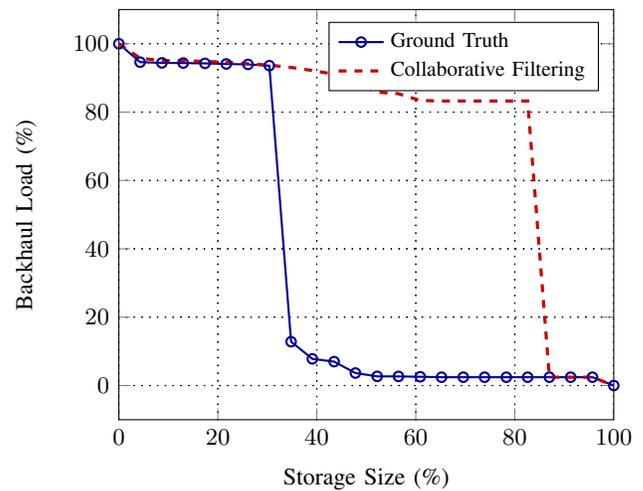

The evolution of users' request satisfaction with respect to the storage size is given Fig. \ref{fig:satisfaction}. The storage size is given in terms of percentage where $100\%$ of storage size represents the sum of all size of contents in the catalog ($17.7$ GByte). From zero storage ($0\%)$ to full storage ($100\%$), we can seen that the users' request satisfaction increases monotonically and goes up to $100\%$, both in ground truth and collaborative filtering approaches. However, there is a performance gap between the ground truth and \ac{CF} until $87\%$ of storage size, which is due to the estimation errors. For instance, with $40\%$ of storage size, the ground truth achieves $92\%$ of satisfaction whereas the \ac{CF} has value of $69\%$.

The evolution of backhaul load/usage with respect to the storage size of \glspl{SBS} is given in Fig. \ref{fig:offloading}. As the storage size of \glspl{SBS} increases, we see that both approaches reduces backhaul usage (namely higher offloading gains). For example, with $87\%$ of storage size for caching, both approaches offload $98\%$ of backhaul usage. The performance of ground truth is evidently higher than the \ac{CF} as all of the available information is taken into consideration for caching. We also note that there is a dramatical decrease of backhaul usage in both approaches after a specific storage size. In fact, most of the previous works on caching assume a content catalog with identical content sizes. In our case, we are dealing with real traces in the numerical setup where the size of contents differs from content to content, as discussed in the previous section (see Fig. \ref{fig:globalpop-cumsize}). According to this scenario, on the one hand, caching a highly popular content with very small size might not reduce the backhaul usage dramatically. On the other hand,  caching a popular content with very high size can dramatically reduce the backhaul usage. Therefore, as the \ac{CF} approach used here is solely based on content popularity, it fails to capture these content size aspects on the backhaul usage, which in turn results in higher storage requirements to achieve the same performance as in the ground truth. This shows the importance of size distribution of popular contents.

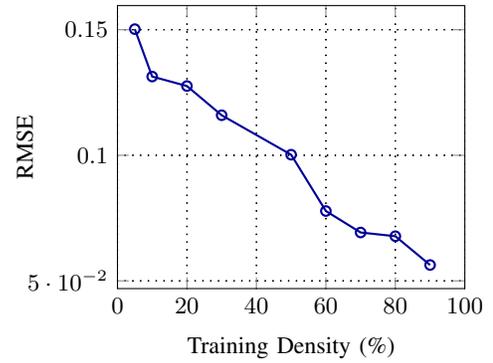
\begin{figure}[!ht]
\centering
\begin{tikzpicture}
	\begin{axis}[
		width=0.70\columnwidth,
 		grid = major,
 		legend cell align=left,
 		xmin=0,xmax=100,	
 		legend style ={legend pos=north east},
 		xlabel={Training Density (\%)},
 		ylabel={RMSE}]

 		\addplot+[blue!60!black, mark=o, mark size=1.8, thick]
 				  table [col sep=comma] {\string"results-rmse-satisfactionRatio.csv"};
		 		  		
	\end{axis}
\end{tikzpicture}
\caption{Evolution of \ac{RMSE} with respect to the training density.}
\label{fig:rmse}
\end{figure}

We have so far compared the performance gains of these approaches with $10\%$ of rating density in \ac{CF}. In fact, as the rating density of \ac{CF} for training increases, we expect to have less estimation error, thus resulting closer satisfaction gains to the ground truth. To show this, the change of \ac{RMSE} with respect to the training rating density is given in Fig. \ref{fig:rmse}. Therein, we define the error as the root-mean-square of difference between users' content satisfaction of the ground truth and \ac{CF} approaches over all possible storage sizes. Clearly, as observed in Fig. \ref{fig:rmse}, the performance of \ac{CF} is improved by increasing the rating density, thus confirming our intuitions.
%
\section{Conclusions}
\label{sec:conclusions}
In this work, we have studied a proactive caching approach for $5$G wireless networks by exploiting large amount of available data and employing machine learning tools. In particular, an experimental setup for data collection/extraction process has been demonstrated on a big data platform and machine learning tools (\ac{CF} in particular) have been applied to predict the content popularity distribution. Depending on the rating density and storage size, the numerical results showed that several caching gains are possible in terms of users' request satisfactions and backhaul offloadings. An interesting future direction of this work is to conduct a more \emph{detailed characterization of the traffic} which captures different spatio-temporal content access patterns. In order to estimate the content access patterns for cache decision, the development of \emph{novel machine learning algorithms} is yet another interesting direction. Finally, design of new  \emph{deterministic/randomized cache decision algorithms} are required and should not be purely based on content popularity and storing most popular contents, so that higher backhaul offloading can be achieved while satisfying users' requests.
\bibliographystyle{IEEEtran}
\bibliography{references}
\end{document}